\documentclass[12pt]{article}
\usepackage[english]{babel}
\usepackage{epsf,epsfig}
\usepackage{amsmath,amsfonts,amssymb}
\usepackage{latexsym}
\usepackage{graphicx,color}
\usepackage{cite}
\usepackage{verbatim}
\usepackage{amsthm}
\usepackage{amsbsy}
\usepackage{hyperref}
\definecolor{link}{rgb}{0.9,0.1,0.1}
\hypersetup{colorlinks=true,linkcolor=link,citecolor=link,urlcolor=link,linktocpage}
\newcommand{\be}{\begin{equation}}
\newcommand{\ee}{\end{equation}}
\newcommand{\ben}{\begin{eqnarray}\displaystyle}
\newcommand{\een}{\end{eqnarray}}

\newcommand{\bea}[1]{\begin{eqnarray}\label{#1} }
\newcommand{\eea}{\end{eqnarray}}

\newcommand{\refb}[1]{(\ref{#1})}

\def\boxempty{{\,\lower0.9pt\vbox{\hrule \hbox{\vrule height 0.25 cm
\hskip 0.25 cm \vrule height 0.25 cm}\hrule}\,}}

\begin{document}
\begin{titlepage}
\thispagestyle{empty}

\title{
{\Large\bf Weak Coupling Expansion of Yang-Mills Theory}\\ 
{\Large\bf  on Recursive Infinite Genus Surfaces
}}

\bigskip\bigskip\bigskip

\author{{\bf Debashis Ghoshal}${}^1$\thanks{{\tt dghoshal@mail.jnu.ac.in}} , 
              {\bf Camillo Imbimbo}${}^2$\thanks{{\tt camillo.imbimbo@ge.infn.it}}\\
              {\bf and Dushyant Kumar}$^1$\thanks{{\tt sehrawat.dushyant@gmail.com}}\\
\hfill\\              
${}^1${\it School of Physical Sciences, Jawaharlal Nehru University}\\
{\it New Delhi 110067, India}\\
\hfill\\
${}^2${\it Dipartimento di Fisica, Universit\`{a} degli studi di Genova}\\
and\\
{\it INFN, Sezione di Genova} \\
{\it  Via Dodecaneso, 16146 Genova, Italy}\\
     }

\bigskip\bigskip

\date{%
%
\bigskip\bigskip
\begin{quote}
\centerline{{\bf Abstract}}
{\small
We analyze the partition function of two dimensional Yang-Mills theory on a family of
surfaces of infinite genus. These surfaces have a recursive structure, which was used
by one of us to compute the partition function that results in a generalized  Migdal formula.
In this paper we study  the  `small area' (weak coupling) expansion of the partition function,
by exploiting the fact that the generalized Migdal formula is analytic in  the (complexification 
of the) Euler characteristic. The structure of the perturbative part of the weak coupling expansion 
suggests that the moduli space of flat connections (of the SU(2) and SO(3) theories) on these 
infinite genus surfaces are well defined, perhaps in an appropriate regularization. 
}
\end{quote}
}

\bigskip


\end{titlepage}
\maketitle\vfill \eject

\tableofcontents

\section{Introduction}\label{sec:Introd}
Yang-Mills gauge field theory in two dimensions (2dYM) has been studied for a long time for its 
rich mathematical and physical content\cite{'tHooft:1974hx,Migdal,Witten:1991we,Witten:1992xu}.   
It also serves as a testing ground for new ideas. 2dYM and its quantum deformation have been 
found to be related to statistical systems, topological strings and even black holes
\cite{Cordes:1994sd,Vafa:2004qa,deHaro:2004id,Ooguri:2004zv,Aganagic:2004js,Caporaso:2006kk}.
Since gauge fields in two dimensions do not have any local propagating degrees of freedom, one 
would surmise that 2dYM is a topological quantum field theory, one that depends only on the global 
properties of two dimensional space-time.  As a matter of fact it turns out that the theory is only `almost' 
topological. Indeed, the lattice action on a triangulation of the surface leads to an exact expression
\cite{Migdal} for the partition function (Euclidean path integral) that depends on the area of the surface, 
as well as its Euler characteristics. The area of the surface is the only `non-topological' parameter, 
and it is really the gauge coupling constant.

The topological content of 2dYM  was unravelled by Witten \cite{Witten:1991we,Witten:1992xu} (see 
also the extensive review\cite{Cordes:1994fc}), who reconsidered the model from the point of view of topological 
quantum field theories, which were developed at that time. The observation that the theory is indeed
topological in the zero-coupling limit, led to the identification of the partition function of physical YM 
at non-zero coupling with the generating function of the correlators of certain BRST invariant observables 
of the topological YM. The topological supersymmetry underlying 2dYM allows the partition function to 
{\em localize}\footnote{Indeed 2dYM was the first quantum field  theory in the context of which the methods of 
non-abelian  localization for quantum field theories were developed and tested. More recently the idea of localization 
has lead to enormous progress in our understanding of  (supersymmetric) field theories in diverse dimensions. 
An incomplete set of references are \cite{Witten:1992xu,Blau:1995rs,Witten:1994ev,Lozano:1999us,%
Nekrasov:2002qd,Beasley:2005vf,Giombi:2009ms,Pestun:2009nn,Giombi:2009ds}.} around the space of 
inequivalent flat gauge connections. The moduli space $\mathcal{M}_g$ of flat gauge connections, depends 
on the gauge group $G$ as well as the genus $g$ of the two dimensional surface (Euclidean space-time). 
The correspondence (between the topological and physical theory) means that the coefficients of the 
perturbative expansion of the partition function of physical 2dYM are related to the integrals of certain 
cohomology classes on $\mathcal{M}_g$. These integrals are very hard to compute directly for generic $g$ 
(and $G$), since the dimension  $\mathcal{M}_g$ grows with the genus and its topology also becomes more 
intricate at higher genera. For this reason, the computation of these numbers using mathematical tools are 
limited. (Some relevant mathematical results on the moduli spaces are to be found in Refs.
\cite{NaraSesh,AtiyBott,Thaddeus,King:1994nt,Fock:1998nu,Liu:2003qe,Sengupta:2007wh}.)
The correspondence of Ref.\cite{Witten:1992xu} allows for a computation using simpler `physicists' method' 
for any $g$ and generic gauge groups. The moduli spaces $\mathcal{M}_g$ are also of considerable physical 
importance: as the phase space of Chern-Simons theory in three dimensions and in the description of WZW 
models, which form an important class of two dimensional conformal field theories. More generally these 
moduli spaces are low dimensional analogues of the moduli space of instantons of four 
dimensional Yang-Mills theory, the geometry of which is crucial for the understanding of the dynamics of 
(supersymmetric) gauge theories and their dualities in four dimensions.

Recently, one of us studied 2dYM on a class of surfaces of infinite genus. In particular, the recursive structure 
of these surfaces was exploited to arrive at a formula for the partition function\cite{Kumar:2014jpa}. We shall refer to 
these as {\em Richards surfaces} after Richards, who classified surfaces of infinite characteristics\cite{Richards}. 
Strictly speaking, however, our surfaces are not among those in Ref.\cite{Richards}, which are surfaces with `ideal 
boundaries' at infinity, characterized by subsets of the Cantor set. On the other hand, the surfaces we consider, 
obtained from different compactifications, parametrized by two integers, of a surface in \cite{Richards}, have 
singularities in place of the `ideal boundaries'. The partition function of 2dYM on these surfaces turns out to bear 
a surprising resemblance to the Migdal formula. Indeed, the Euler characteristic is replaced by a {\em rational} 
number, which may be thought of as the formal Euler characteristic of the recursive surface. 

In the light of what we have reviewed above, one would expect that the perturbative expansion of the 
generalized Migdal formula of  \cite{Kumar:2014jpa} encodes integrals of appropriate forms on the moduli space 
$\mathcal{M}\left(\bar{\mathcal{S}}_{g,p}\right)$ of flat connections on the  Richards surfaces 
$\bar{\mathcal{S}}_{g,p}$ parametrized by the integers $g$ and $p$. The mathematical properties of this  
moduli space are, as yet, largely unknown. However we expect it to be infinite-dimensional. Therefore we can 
anticipate that the {\em weak coupling} expansion of the generalized Migdal formula of  \cite{Kumar:2014jpa} 
contains exact information about (an infinite number of) integrals of appropriate forms on the infinite dimensional 
moduli spaces. The correspondence between the topological and physical 2dYM on Richard surfaces may, therefore, 
provide the first known example of a topological quantum theory which localizes on an infinite-dimensional moduli 
space in an {\it exactly computable} way.  Moreover one expects  $\mathcal{M}\left(\bar{\mathcal{S}}_{g,p}\right)$ to
be  relevant for higher-dimensional gauge theories in the same way as the finite genus $\mathcal{M}_g$ is. An 
additional motivation to study 2dYM on Richard surfaces is that analytical results in this topic  may give some 
insight to the old ideas\cite{FS, LaThermo} concerning the significance of 2d CFT  on infinite genus surfaces for 
studying non-perturbative aspects of string theory.

In order to extract the intersection numbers on the moduli spaces of flat connections, one has to go from {\em strong 
coupling}, which is the regime of validity of the Migdal formula, to {\em weak coupling}, the regime where the 
correspondence with topological 2dYM holds. Even in the finite genus case this is not trivial to do.  This was achieved 
by a trick in Refs.\cite{Witten:1991we,Witten:1992xu} by considering a certain derivative of Migdal's partition function 
with respect to the coupling constant. This results in a theta function which can be Poisson resummed to generate the 
weak coupling expansion. It turns out that the {\it perturbative} part of the weak coupling expansion stops at a {\it finite} 
order, which depends on $g$. This is a reflection of the finite dimensionality of $\mathcal{M}_g$. The weak coupling 
expansion also contains non-perturbative contributions which are not related to  $\mathcal{M}_g$ but represent the 
contributions from the moduli spaces of {\it non-flat} instantons.

Since the trick of \cite{Witten:1991we,Witten:1992xu} does not work for the Richards surfaces, we adopt a different 
strategy. We observe that the generalized Migdal formula derived in \cite{Kumar:2014jpa} can be analytically 
continued to complex values of the generalized Euler characteristics. In a suitable region of the complex plane (of 
the generalized Euler charateristics) the  partition function is a sum over integer values of a function which admits 
a  Fourier transform.  This allows for a Poisson resummation of the generalized Migdal formula even for the Richards 
case. We perform this explicitly for  the SO(3) and SU(2) theories --- although the strategy may be applied to generic 
gauge groups --- and derive explicit formulas for the invariants, which should be identified with the integrals on the 
infinite dimensional moduli spaces $\mathcal{M}\left(\bar{\mathcal{S}}_{g,p}\right)$, which are yet to be understood 
mathematically. 

Our weak-coupling formula for YM on Richard surfaces also includes all the exponentially small, non-perturbative 
contributions, which are not related to $\mathcal{M}_g$, but represent the contributions to the partition function of 
moduli spaces of  {\it non-flat} instantons. A detailed analysis of  these terms is expected to shed light on the
resurgence structure in a quantum field theory, however, we leave it as a project for the future.

In the following, we review the construction of a Richards surface and the generalized Migdal formula for 
the partition function of 2dYM on it\cite{Kumar:2014jpa}. We also briefly recapitulate the results from Refs.
\cite{Witten:1991we,Witten:1992xu} on (finite genus) Riemann surfaces. We then perform the weak coupling 
expansion of the partition function on Richards surface. We find that, in this limit, the partition function can be 
separated into a part which is analytic and another which vanishes exponentially. (For the SU(2) case, there are 
also non-analytic terms attributable to singularities in the moduli space.) This parallels the case of surfaces of finite 
characteristics for which the analytic part is a polynomial. Indeed in such cases, the formal Euler characteristic 
coincides with the Euler characteristic, and our results coincide with those in Refs.\cite{Witten:1991we,Witten:1992xu}, 
thanks to non-trivial identities of zeta functions. We will also compute the partition function (of the SU(2) theory for 
definiteness) with a fixed (nontrivial) holonomy along a boundary of the surface. This results in a generating function 
with two parameters.

\section {Richards surfaces}\label{sec:RichCons}
In Ref.\cite{Richards}, Richards gave a classification of noncompact surfaces, which have 
`ideal boundaries' at asymptotic infinity labelled by a subset of the Cantor set. Recall that a 
Cantor set may be constructed by removing every alternate $\ell$ intervals out of the $(2\ell+1)$ 
segments of $\left[0,1\right]$, for any positive integer $\ell$. This process is then repeated for every 
remaining intervals of the first set, and so on ad infinitum. The result is a totally disconnected space. 

\begin{figure}[ht]
\begin{center}
\includegraphics[scale=.25]{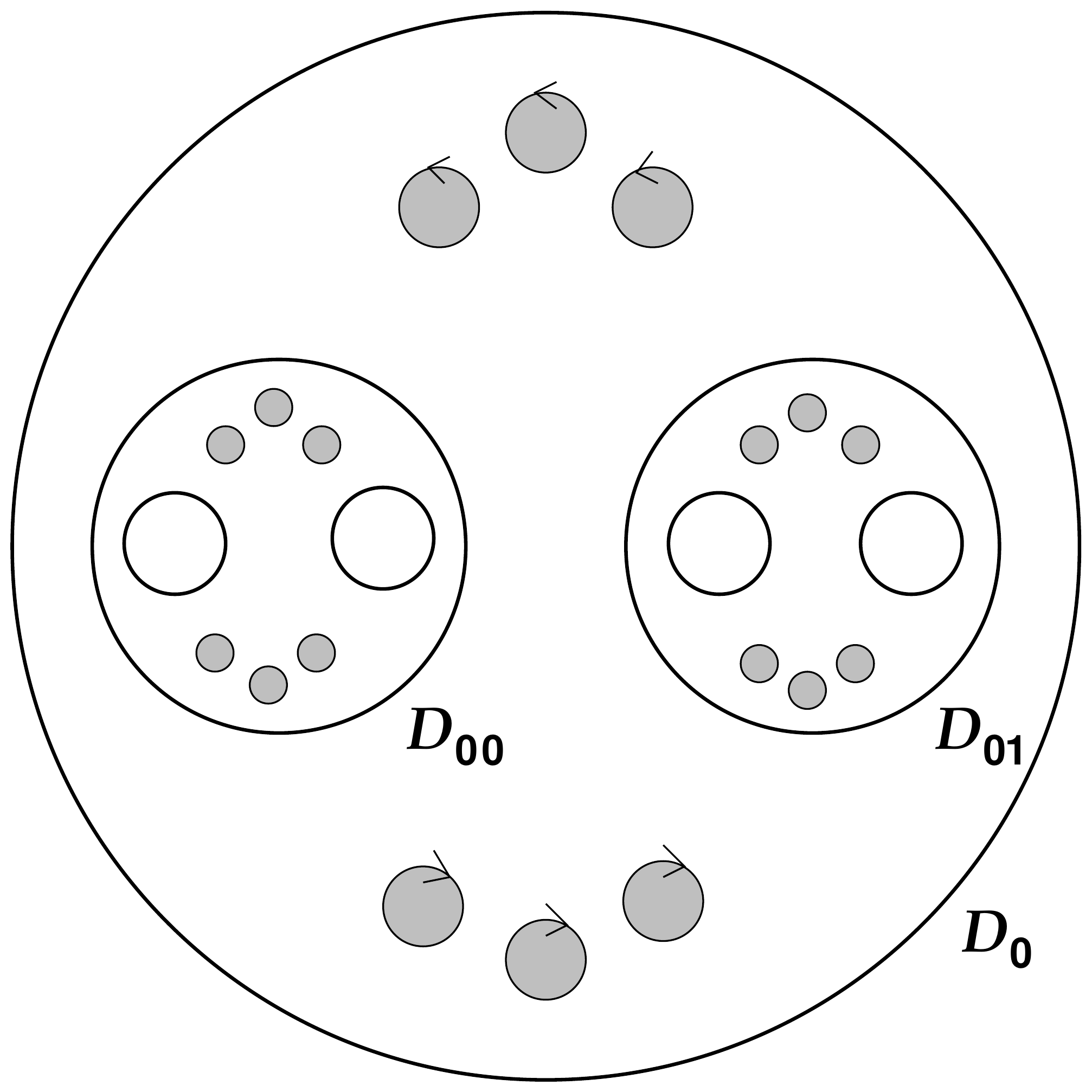}
\hspace*{12pt}
\includegraphics[scale=.85]{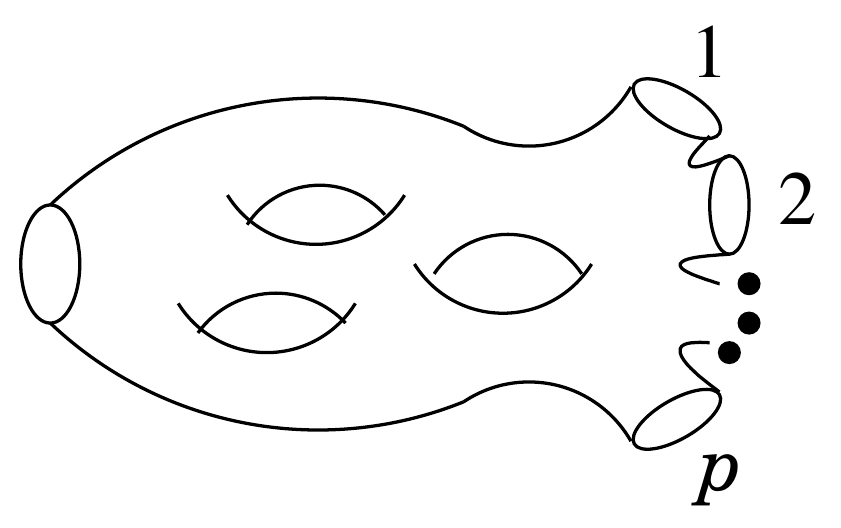}
\caption{{\small
Left: Iterative construction of a Richards surface starting with the disc ${\cal D}_0$. In this figure
$p=2$ and $g=3$.
Right: The basic building block of a Richards surface with $g=3$ and an arbitrary number of
branchings.
}}
\label{fig:RichardsSurf}
\end{center}
\end{figure}

Motivated by the arguments in \cite{Richards}, one of us considered a family of surfaces, which we
refer to as {\em Richards surfaces}. These may be thought of as compactifications of a surface in
Richards' classification. The points in the set of ideal boundary, homeomorphic to the Cantor set, are 
singular points of these surfaces. Let us recall this construction. We start with a closed disc 
${\cal D}_0$ containing the interval $[0,1]$, which in turn, contains the Cantor set. To be specific, 
we consider, $\ell=1$, for which we can take the discs ${\cal D}_{00}$ and ${\cal D}_{01}$ which 
contain the intervals $\left[0,\frac{1}{3}\right]$ and $\left[\frac{2}{3},1\right]$, respectively, and do 
not overlap. Let us attach $g$ handles to the complement of ${\cal D}_{00}\cup{\cal D}_{01}$ in 
${\cal D}_{0}$.  Equivalently, we remove $2g$ disjoint discs in the complement and identify the 
boundaries of each pair preserving orientation. In the next step, perform similar operations in
${\cal D}_{00}$ and ${\cal D}_{01}$, and so on. The limit of this iterative process is an infinite genus 
surface which is made up of segments with $g$ handles and $p=2$ boundaries, to each of which is 
attached another similar segment. We refer to it as a Richards surface with parameters $p$ and 
$g$ (see Fig.~\ref{fig:RichardsSurf}). 

The limiting sets ${\cal D}_{000\cdots}$, ${\cal D}_{001\cdots}$, ${\cal D}_{010\cdots}$, etc.\ contain
the singular points of this surface. In Ref.\cite{Richards}, where the points of the Cantor set are removed
from ${\cal D}_0$ from the beginning, these limiting sets would have been the `ideal boundaries' instead. 
The noncompact surface ${\cal S}$ obtained by the latter procedure can be shown to be independent of 
the choices of integers $g$ and $p$, and therefore, results in a unique (upto homeomorphism) surface. 
Therefore, our surfaces $\bar{\cal S}_{g,p}$ are to be thought of as compactifications of it.

In Ref.\cite{Kumar:2014jpa} the recursive nature of these surfaces was used to compute the partition 
function of 2dYM for gauge group $G$: for a Richards surface $\bar{\Sigma}_{g,p}$ given by the 
parameters $g$ (for the numbers of handles attached to each building block) and $p$ (for the number 
of branchings at each step) is
\begin{equation}
Z^{\left(g,p\right)}_\infty\left(\epsilon;U_0\right)\, = \, \sum_R \left(\mathrm{dim}\, R\right)^{1 + 
\frac{2g}{p-1}}\, e^{- 4\pi^2 \epsilon\, C_2({R})}\; \chi_R(U_0), \label{pfYMRichards}
\end{equation}
where the sum is over all irreducible representations $R$ of the Lie group $G$, $C_{2}(R)$ 
and $\chi(R)$ are, respectively, the second Casimir operator and the character corresponding 
to $R$ and the dimensionless parameter $\epsilon = a\tilde{g}^2/4\pi^2$ is given in terms of the 
(regularized) area $a$ of the surface and the coupling constant $\tilde{g}^2$ of the gauge theory. 
We have also put a holonomy $U$ along the boundary of the surface. (This is the boundary of the
initial disc ${\cal D}_0$.) 
This formula is not applicable to $p=1$, which results in a surface known as the Loch Ness 
monster. However, it is consistent with the known results for the special cases (i) $g=0$ (a disc) 
and (ii) $p=0$ (a surface of finite genus $g$). 

The partition function of Richards surface is remarkably similar to the Migdal formula for surfaces
with finite genus\cite{Migdal}. The main difference is that the power of dim $R$ is fractional. In fact,
being the formal sum of the characteristics of its buildings blocks, it may be thought of as the Euler
characteristic of the Richards surface.

We see from the partition function \refb{pfYMRichards} above, that Richards surfaces with different 
parameters $g$ and $p$ have generically different partition functions. Hence we may conclude (at 
the level of rigour used by physicists) that they are topologically inequivalent. 

\section{Yang-Mills theory on Riemann surfaces}\label{sec:YMfinite}
Let us recall the analysis of the partition function of 2dYM in Refs.\cite{Witten:1991we,Witten:1992xu}
to extract intersection numbers of cycles in the moduli space of flat connections on a Riemann
surface. We start with the partition function of a surface of genus $g$ and the number of boundary 
components $p$:
\begin{equation}
Z_{\left(g,p\right)}(\epsilon; U_{1},\cdots,U_{p}) = \displaystyle\sum_{R}(\dim R)^{2-2g-p}\; 
e^{- 4\pi^2 \epsilon\, C_{2}\left( R\right)}\;\chi_R(U_{1}) \cdots\chi_R(U_{p}),
\label{MigdalForm} 
\end{equation}
where $\epsilon = \displaystyle{\frac{a\tilde{g}^2}{4\pi^2}}$ is a dimensionless parameter defined earlier. 
The holonomies $U_1,\cdots, U_p$ are associated with the $p$ boundaries, and the second Casimir 
invariant $C_2\left(R\right)$ corresponds to the representation $R$. This partition function comes 
from integrating the exponential of the Yang-Mills action functional over the space of gauge
inequivalent connections with specified holonomies along the given boundaries. 

In the limit of $\epsilon\to 0$, the most dominant contribution to the partition function is from the integration 
over the space of flat connections. Thus,  in this limit, one can expect the partition function to contain some 
topological information about the moduli space. This indeed is the case. The small area expansion of the 
partition function around $\epsilon=0$ consists of a polynomial part (also in addition, possibly some terms 
with fractional powers of $\epsilon$), and a part that is exponentially suppressed. The coefficients of the 
polynomial are interpreted as the intersection numbers of the homology cycles on the moduli space of flat 
connections. Let us recall the case of SU(2) and SO(3) groups, for which these computations have been 
performed independently\cite{Thaddeus}.

\subsection{SU(2) theory}\label{subsec:YMfiniteSU}
Only a trivial principal $SU(2)$ bundle is possible for a closed surface $\Sigma$ of genus $g$. Let us 
denote it as $E$. The exponential of the Yang-Mills action integrated (with respect to the symplectic 
measure) over the space of connections on $E$ gives the following partition function 
\begin{equation}
Z_g(\epsilon)={\displaystyle \frac{1}{(2\pi^{2})^{g-1}}\sum_{n=1}^{\infty}}n^{2-2g}\:
\exp\left(- \,\epsilon\,\pi^2 n^{2}\right), 
\label{eq:1}
\end{equation}
This is the Migdal formula for SU(2), normalized by the volume of the gauge group as in 
Refs.\cite{Witten:1991we,Witten:1992xu}.  

Let $\mathcal{M}$ denote the moduli space of (gauge equivalence classes of) flat SU(2) connections 
on $E$. The partition function \refb{eq:1} for $g\geq1$ is related to the integration of the closed differential 
forms $\omega$ and $\Theta$, of degree 2 and 4 respectively, on $\mathcal{M}$. Of these, $\omega$ is 
the pull-back on $\mathcal{M}$ of the natural symplectic form 
\begin{equation}
\omega(\delta A , \delta A) = \int_\Sigma \mathrm{Tr}\,   \delta A\wedge \delta A
\end{equation}
which exists on the space of SU(2) connections on $\Sigma$. On the other hand, as explained in 
\cite{Witten:1992xu}, $\Theta$ has a natural interpretation in the two dimensional topological Yang-Mills 
theory. The topological YM gauge multiplet contains, in addition to the connection field $A_\mu(x)\,dx^\mu$ 
and its fermionic super-partner  $\psi_\mu(x)\, dx^\mu$ of ghost number 1, a scalar field  $\phi(x)$ of ghost 
number 2, all in the adjoint representation of  the gauge group. The cohomology class of $\Theta$ is 
associated to the  BRST invariant local observable $\mathrm{Tr} \,\phi^2(x)$. It turns out that for fixed 
non-negative integers $k\leq\frac{1}{4}\mbox{dim}\mathcal{M}$ and 
$s=\frac{1}{2}\mbox{dim}\mathcal{M}-k$,
\begin{equation}
\frac{1}{(s-k)!} \displaystyle\int_{\mathcal{M}}\omega^{s-k}\wedge\Theta^{k}=
2\,\frac{d^{k}}{d\epsilon^{k}} Z_g(\epsilon)\bigg|_{\epsilon=0},\label{eq:2}
\end{equation}
whenever the LHS and RHS of this equation make sense. The factor of $2$ on the RHS is the number 
of elements in the centre of SU(2). The above equation can be written in a compact form as 
\begin{equation}
\displaystyle\int_{\mathcal{M}}\,\exp(\omega+\epsilon\, \Theta)=
2\,Z_g(\epsilon)\quad \text{(modulo exponentially small terms)}.\label{eq:3a}
\end{equation}

The expansion in Migdal formula Eq.\refb{eq:1} is useful for $\epsilon>0$ large. To extract the 
$\epsilon\to0^+$ behaviour of the partition function and make contact with the topological numbers in 
Eq.  \refb{eq:2} one cannot simply expand the exponentials in the series and take the limit, since this
would result in a divergent expression. The trick used in Refs.\cite{Witten:1991we,Witten:1992xu} is to 
consider the derivative 
\begin{eqnarray}
\frac{d^{g-1}}{d\epsilon^{g-1}} Z_g(\epsilon) &=& \frac{(-1)^g}{2^{g-1}} \,{\displaystyle \sum_{n=1}^{\infty}}
\exp\left(- \,\epsilon\,\pi^2 n^{2}\right)\nonumber\\
&=& \frac{(-1)^{g-1}}{2^{g}}\left[ \,{\displaystyle \sum_{n=-\infty}^{\infty}}
\exp\left(- \,\epsilon\, \pi^2 n^{2}\right) - 1\right], \label{eq:1bis}
\end{eqnarray}
and to make use of the Poisson resummation formula
\begin{equation}
{\displaystyle \sum_{n=-\infty}^{\infty}} f(n)= {\displaystyle \sum_{k=-\infty}^{\infty}}\, \tilde{f}(k)
\label{eq:1tris}
\end{equation}
(where $ \tilde{f}$ is the Fourier transform of $f$)  to obtain
\begin{eqnarray}
\frac{d^{g-1}}{d\epsilon^{g-1}} Z_g(\epsilon) &=& \frac{(-1)^{g-1}}{2^{g}} \, \left[\frac{1}{\sqrt{\pi\,\epsilon}}\,  
{\displaystyle \sum_{k=-\infty}^{\infty}}\exp\left(- \frac{\pi^2}{\epsilon}\, k^{2}\right) - 1\right] \nonumber\\
&=& \frac{(-1)^{g-1}}{2^g} \,\left(\frac{1}{\sqrt{\pi\,\epsilon}} - 1\right) + \text{exponentially small terms}.
\label{eq:1quater}
\end{eqnarray}
This determines $Z_g(\epsilon)$ up to a polynomial $P^{(g-2)}(\epsilon)$ in $\epsilon$ of degree $g-2$ 
(for $g\ge 2$). 

It is the polynomial which contains the information about the intersection numbers \refb{eq:2}.
In order to determine the coefficients of this polynomial, one can evaluate the first $k$ derivatives 
of the Migdal formula at $\epsilon=0$
\begin{equation}
\left.\frac{d^{k}}{d\epsilon^{k}} Z_g(\epsilon)\right|_{\epsilon=0}=
\frac{(-1)^k}{2^{g-1}\, \pi^{2\,(g-1-k)}} \,{\displaystyle \sum_{n=1}^{\infty}}\,\frac{1}{n^{2(g-1-k)}}
\label{eq:1penta}\qquad \text{for}\quad k< g-1.
\end{equation}
in which the series on the RHS are convergent for $k<g-1$. In this way, one finds that
\begin{equation}
Z_g(\epsilon) = \displaystyle\sum_{k=0}^{g-2} a_k\,\epsilon^k + 
a_{g-\frac{3}{2}}\,\epsilon^{g-\frac{3}{2}} + a_{g-1}\,\epsilon^{g-1} +
\;\text{exponentially small terms},\label{eq:mj}
\end{equation}
where the coefficients are
\begin{eqnarray}
a_k &=& \frac{(-1)^k}{k!\, 2^{g-1}\, \pi^{2g-2-2k}}\;\zeta\left(2g-2-2k\right)\:\:
\text{ for } k=0,1,\cdots,g-2,\label{coeff-poly}\\
a_{g-\frac{3}{2}} &=& \frac{ (-1)^{g-1}}{2\, \sqrt{\pi}\,\left(2g-3\right)!!}\:\:
\text{ and }\: 
a_{g-1} \:=\: \frac{ (-1)^{g}}{2^{g}(g-1)!}.\nonumber
\end{eqnarray}
Note that  the coefficient of the nonanalytic term, $a_{g-\frac{3}{2}}$, is nonzero, which is a manifestation 
of the fact that the moduli space of SU(2) flat connections is  singular. This is due to that fact that SU(2) has 
a nontrivial centre and hence the action of SU(2) gauge  group on the space of flat connections is not free.  
One can read off the intersection numbers on $ \mathcal{M}$, given in Eq.\refb{eq:2}, from Eqs.\refb{eq:mj} 
and \refb{coeff-poly}. (We note parenthetically that a factor of $-1$ is missing from Eqs.(4.52) and (4.53) in 
\cite{Witten:1992xu} due to a typographical error. The corresponding expressions are 
Eqs.\refb{eq:1bis} and \refb{eq:1quater} above.)

\subsection{SO(3) theory}
\label{sec:YMfiniteSO}
The calculations for SO(3) are similar to that for SU(2) except for a few important differences.
First, for a closed surface $\Sigma$ of genus $g\geq1$, two topologically inequivalent SO(3) 
principal bundles $E'(1)$ and $E'(-1)$ are possible. Of these, $E'(1)$ is trivial and lifts to an SU(2) 
bundle on $\Sigma$. On the other hand $E'(-1)$ is nontrivial and can be lifted to an SU(2) bundle 
$E_p$ only on the complement $\Sigma-\{p\}$ of a point $p\in\Sigma$. If we choose a flat SU(2) 
connection $A$ on $E_p$ then the holonomy of this connection about $p$ would be $-1$, 
independent of the choice of $A$ and $p$. 

Let $u=\pm 1$, then the partition function of SO(3) Yang-Mills theory on the space of connections 
on $E'(u)$ is given by 
\begin{equation}
\widetilde{Z}_g(\epsilon; u) = \frac{1}{2(8\pi^{2})^{g-1}} \displaystyle\sum_{n=1}^{\infty}
n^{2-2g}\; \lambda_{n}(u^{-1})\; \exp\left(-\,\epsilon\, \pi^2 n^{2}\right),\label{eq:3}
\end{equation}
where $\lambda_{n}(u^{-1})=\chi_{n}(u^{-1})/n$ in which $\chi_{n}(u^{-1})$ is the trace of $u^{-1}$ 
in the representation of dimension $n$. The sum is over all the representations of SU(2) instead of 
only those of SO(3). However, the sum $\widetilde{Z}_g(\epsilon) = 
\widetilde{Z}_g(\epsilon;\;1) + \widetilde{Z}_g(\epsilon;\;-1)$ (which, in principle, is the 
full SO(3) partition function) will have contribution only from the irreducible representations of 
SO(3).
 
Secondly, let $\mathcal{M}'(-1)$ denote the moduli space of (gauge equivalent classes of) flat 
SO(3) connections\footnote{The case of $\mathcal{M}'(1)$ is exactly same as that for SU(2).} for 
the bundle $E'(-1)$. The partition function Eq.\refb{eq:3} for $g\geq1$ is related to the integration 
of the differential forms $\omega$ and $\Theta$ on $\mathcal{M}'(-1)$ in the following 
way\cite{Witten:1992xu}. For fixed non-negative integers 
$k\leq\frac{1}{4}\mbox{dim}\mathcal{M}'(-1)$ and $s=\frac{1}{2}\mbox{dim}\mathcal{M}'(-1)-k$, 
\begin{equation}
\frac{1}{(s-k)!} \displaystyle\int_{\mathcal{M}'(-1)} \omega^{s-k}\wedge\Theta^k =
\left.\frac{d^{k}}{d\epsilon^{k}} \widetilde{Z}_g(\epsilon;-1)\right|_{\epsilon=0},\label{eq:4}
\end{equation}
where there is not any factor of $2$ this time because the centre of SO(3) is trivial. The numbers 
$\left.\displaystyle{\frac{d^{k}}{d\epsilon^{k}}} \widetilde{Z}(\Sigma,\epsilon;-1)\right|_{\epsilon=0}$ 
can be calculated in exactly the same way as for SU(2): for $k\leq (g-2)$ directly from Eq.\refb{eq:3}, 
and for $k\geq (g-1)$ by using Poisson resummation. We have
\begin{equation}
Z_g(\epsilon) = \displaystyle\sum_{k=0}^{g-1} a_k\,\epsilon^k + 
\;\text{exponentially small terms},\label{eq:mj2}
\end{equation}
where the coefficients are
\begin{equation}
a_k = (-1)^k\; \frac{(1-2^{2g-3-2k})\:
\zeta(2g-2-2k)}{k!\; 2^{3g-3}\; (2\pi)^{2g-2-2k}}.\label{coeff-poly-so3}\\
\end{equation}
Notice that the RHS does not have any term with fractional powers of  $\epsilon$ as $\mathcal{M}'(-1)$ 
is nonsingular, and $a_0$ is its symplectic volume.

\section {Analysis of 2dYM on Richards surfaces}\label{sec:RichYM}
Richards surfaces  have a self-similar structure which allows one to compute the partition function of 
Yang-Mills theory on them. The partition function for a surface $\bar{\cal S}_{g,p}$ is given in 
Eq.\refb{pfYMRichards}. As we have pointed out earlier, the only difference 
from the partition function of finite genus surfaces, Eq.\refb{MigdalForm}, is that the power of dim $R$ 
is in general a positive fraction (which may be thought of as the formal Euler characteristic). In the following 
we will work with the assumption that the partition function of 2dYM on Richards surfaces would allow
for an interpretation as a generating function of intersection numbers on moduli space of flat connections
of  principal $G$-bundle on Richards surface, exactly analogous to the finite genus surfaces. As in the 
previous section, we will consider the cases of SU(2) and SO(3).  

\subsection{Gauge group SU(2)}
The extension of the SU(2)  Migdal formula to the infinite genus Richards surfaces\cite{Kumar:2014jpa}
takes the form\footnote{We have ignored the equivalent of the normalization factor 
$(2\pi^2)^{g-1}$ in Eq.\refb{eq:1}, as the exact geometric analogue of it is not yet understood. The
same applies to the Eqs.\refb{eq:3} and \refb{eq:12} for SO(3), which has an additional factor of 2.
\label{ff_volGgroup} } 
\begin{equation}
Z(\alpha;\epsilon) = \displaystyle\sum_{n=1}^{\infty} n^{\alpha}\, \exp\left(-\,\epsilon\,\pi^2n^{2}\right).
\label{eq:5}
\end{equation}
where 
\begin{equation}
\alpha=2+\frac{2g}{p-1}\label{alpha}
\end{equation}
is now a rational number.
It is reasonable to suppose that formula \refb{eq:5} encodes some kind of topological information by means 
of an equation analogous to Eq. \refb{eq:3a}, involving integration of classes on the infinite dimensional 
moduli space $\mathcal{M}\left(\bar{\mathcal{S}}_{g,p}\right)$ of flat connections on the  Richards surfaces 
with fixed $g$ and $p$.
It would be interesting therefore to determine the behaviour of the function in Eq.\refb{eq:5} near $\epsilon=0$.
To obtain this, obviously we cannot exploit the same trick of Refs.\cite{Witten:1991we,Witten:1992xu}, as we 
reviewed in the previous section:  This trick involved taking the $g-1$ derivatives of the partition function with 
respect to $\epsilon$, which  works only when $\alpha= (2-2g)$ is a negative integer. We will therefore adopt 
a different strategy. Let us first of all extend $\alpha$ to the complex plane, and then extend the sum  
\refb{eq:5}  over all integers as
\begin{equation}
Z(\alpha;\epsilon) = \frac{1}{\left(1+e^{i\pi\alpha}\right)}\, \displaystyle\sum_{n=-\infty}^{\infty} 
n^{\alpha}\;\exp\left(-\, \epsilon\,\pi^2 n^{2}\right). \label{eq:8}
\end{equation}
At this point, for $\alpha$ in a suitable region of the complex plane, we can use the Poisson resummation
formula, Eq.\refb{eq:1tris}, with
\begin{equation}
f( x) = x^\alpha\, e^{- \epsilon\, \pi^2 x^2}
\end{equation}
Using the fact that, for $\Re({\alpha})>-1$,
\begin{eqnarray}
{}&{}&\!\!\!\!\!\! \int_{-\infty}^\infty \!\!dx\, x^\alpha\, e^{- \epsilon\, \pi^2 x^2 - 2\pi i k x}\nonumber\\
{}&{}& \qquad =\: 
\frac{\epsilon^{-1-\frac{\alpha }{2}}}{2\pi^{\alpha + 1}}\,
\left[\left(1+(-1)^\alpha\right)\, \sqrt{\epsilon }\, \Gamma\left(\frac{\alpha +1}{2}\right)\, 
{}_1F_1\left(\frac{\alpha +1}{2},\frac{1}{2};-\frac{k^2}{\epsilon }\right)\right.\nonumber\\
{}&{}&\qquad\qquad\qquad\qquad - \,
\left. 2i \left(1-(-1)^{\alpha}\right) \,k\, \Gamma \left(\frac{\alpha}{2}+1\right) \, 
{}_1F_1\left(\frac{\alpha}{2}+1,\frac{3}{2};-\frac{k^2}{\epsilon}\right)\right]
\end{eqnarray}
where ${}_1F_1\left(a,b;z\right)$ is the Kummer confluent hypergeometric function defined by the 
series $\displaystyle\sum_{m=1}^\infty \frac{\left(a\right)_m z^m}{\left(b\right)_m m!}$, in which 
$(a)_m$ and $(b)_m$ are Pochhammer symbols, we obtain 
\begin{equation}
Z(\alpha;\epsilon) = \frac{1}{2\,\pi^{1+\alpha}}\, \Gamma\left(\frac{1+\alpha}{2}\right)\,
{\epsilon}^{-(1+\alpha)/2} \displaystyle\sum_{k=-\infty}^{\infty} {}_1F_1 \left( \frac{1+\alpha}{2}, 
\frac{1}{2}; - \frac{k^2}{\epsilon}\right), \label{eq:9}
\end{equation}
Since we want to know the behaviour of $Z(\epsilon,\alpha)$ near $\epsilon=0$, we use the 
asymptotic expansion 
\begin{equation}
{}_1F_1(a, b; z)\; \stackrel{z\rightarrow -\infty}{\longrightarrow} \; \frac{\Gamma(b)}{\Gamma(b-a)}\, 
\displaystyle\sum_{m=0}^{\infty} \frac{(a)_m\,(a-b+1)_m}{m!}\,(-z)^{-m-a}\; +\; 
{\mathcal O}\left(e^z\right),\label{eq:KummerAsymp}
\end{equation}
of the function ${}_1F_1(z)$ for all $k\neq0$ terms of the series above to write
\begin{eqnarray}
Z(\alpha;\epsilon) &= &  \frac{1}{2\,\pi^{1+\alpha}}\Gamma\left(\frac{1+\alpha}{2}\right) 
{\epsilon}^{-\frac{1+\alpha}{2}}   \label{eq:10}\\
&\!\!+&\!\!\!\!\frac{1}{2\,\pi^{1+\alpha}}\,\Gamma\left(\frac{1+\alpha}{2}\right)
\frac{\sqrt{\pi}}{{\epsilon}^{\frac{1+\alpha}{2}}}\;
\sum_{k\neq0} \frac{1}{\Gamma\left(-\frac{\alpha}{2}\right)} 
\left(\frac{k^2}{\epsilon}\right)^{-\frac{1+\alpha}{2}}
\sum_{m=0}^{\infty} \frac{(\frac{1+\alpha}{2})_m\,(\frac{2+\alpha}{2})_m}{m!} \;
\left(\frac{ k^2}{\epsilon}\right)^{-m} \nonumber\\
&+&  \text{exponentially small terms}. \nonumber
\end{eqnarray}
We can now sum over $k$ to get the final expression for $Z(\epsilon)$ (modulo exponentially small terms):
\begin{eqnarray}
Z(\alpha; \epsilon) &=& \displaystyle{\frac{\Gamma\left(\frac{1+\alpha}{2}\right)}{2\,\pi^{1+
\alpha}{\epsilon}^{\frac{1+ \alpha}{2}}}} + \displaystyle{\frac{\Gamma\left(\frac{1+\alpha}{2}\right)
\sqrt{\pi}}{\pi^{1+\alpha}\Gamma\left(-\frac{\alpha}{2}\right)}} \;
\displaystyle\sum_{m=0}^\infty \displaystyle{\frac{\left(\frac{1+\alpha}{2}\right)_m \left(\frac{2+\alpha}{2}
\right)_m}{ m!}}\,\zeta(2m+1+\alpha)\,{\epsilon}^{m}\nonumber\\
&{}&\quad  +\: \text{exponentially small terms}. 
\label{eq:11}
\end{eqnarray}
The first term in this expression has a fractional negative power of $\epsilon$. We expect this to be related 
to the singularities of the purported moduli space owing to the nontrivial centre of the SU(2), exactly as in 
the case of Riemann surfaces of finite genus for which $\alpha$ is  an even (negative) integer. We will confirm 
this interpretation in the next subsections, where we will show  that the non-analytic terms of this kind are absent 
for both the SO(3) case (in the sector with non-trivial monopole charge) and the case when a non-trivial holonomy 
is inserted at a boundary/puncture on the surface. 

The infinite sum in (\ref{eq:11}) is  the part of the partition function that is {\em analytic} in $\epsilon$. The 
coefficients 
\begin{eqnarray}
 a_m= \frac{\Gamma\left(\frac{1+\alpha}{2}\right)}{\pi^{\frac{1}{2}+\alpha}\Gamma\left(-\frac{\alpha}{2}\right)}\;
 \frac{\left(\frac{1+\alpha}{2}\right)_m \left(\frac{2+\alpha}{2}
\right)_m}{ m!}\;\zeta(2m+1+\alpha),\,\qquad m=0,1,\cdots
\label{eq:11 bis}
\end{eqnarray}
provide a generalization of  the intersection numbers in Eq.( \ref{coeff-poly}), to the Richards Surface. 

The part of $Z(\alpha; \epsilon)$ that is not analytic in $\epsilon$, consists of exponentially small terms. 
These are calculated explicitly in Eq.\refb{eq:9} and can be studied systematically from the asymptotic 
expansion of the confluent hypergeometric function ${}_1F_1$.  They have the structure of instanton 
contributions of the form $e^{-\,\pi^2 k^2/\epsilon}$, just as in the finite genus case.

\subsubsection{Comparison at negative integer \texorpdfstring{$\alpha$}{}}
For the analytic continuation in $\alpha$ to be consistent, one would expect to recover the results of 
Ref.\cite{Witten:1992xu} by setting $\alpha=2-2g$ in the expression above. This is indeed the case.
The Pochhammer symbol $\left(\frac{2+\alpha}{2}\right)_m$ vanishes for all $m\ge g-1$, therefore the 
power series in $\epsilon$ terminates. However, for $m=(g-1)$, the zeta function has a simple pole, 
which compensates for the zero from the Pochhammer symbol and gives a finite contribution, which 
agrees with the coefficient $a_{g-1}$ in Eq.\refb{eq:mj}. Also the  coefficient of 
$\epsilon^{- \frac{1+\alpha}{2}}$ in Eq.\refb{eq:mj}  agrees with the one  in  \refb{coeff-poly}  when
$\alpha=2- 2\,g$.  We are left with a polynomial of degree $(g-2)$ in $\epsilon$ which is identical to the 
corresponding polynomial in Eq.\refb{eq:mj}.  The matching, however, is not trivial, as it involves the 
Riemann functional equation 
\[
\zeta(s) = 2^s\pi^{s-1}\sin\left(\frac{\pi s}{2}\right)\Gamma(1-s)\zeta(1-s)
\]
for the Riemann zeta function, as well as identities for the $\Gamma$-function, the use of which shows 
that 
\begin{eqnarray}
{}&{}& \frac{\sqrt{\pi}\,\Gamma\left(\frac{3}{2}-g\right)}{ \pi^{3- 2\,g}\,\Gamma(g-1)}\;
\sum_{m=0}^{g-2} \frac{\left(\frac{3}{2}-g\right)_m\left(2-g\right)_m}{m!}\,\zeta(2m+3-2g)\epsilon^m
\nonumber\\
{}&{}&\qquad\qquad =\:\:\ \frac{\sqrt{\pi}(-1)^{g-1}}{\pi^{2-2g}\Gamma\left(g-1\right)
\Gamma\left(g-\frac{1}{2}\right)}\;\displaystyle\sum_{m=0}^{g-2} \frac{\Gamma\left(2g-2\right)\,
\zeta\left(2m+3-2g\right)}{m!\,2^{2m}\Gamma\left(2g-2m-2\right)}\,\epsilon^m\nonumber\\
{}&{}&\qquad\qquad =\:\:\sum_{m=0}^{g-2}\frac{(-1)^m\pi^{2m}\zeta(2g-2m-2)}{m!}\epsilon^m,
\label{coeff_match}
\end{eqnarray}
in agreement with Eq.\refb{coeff-poly}. (The coefficients match precisely when the volume factor mentioned
in footnote \ref{ff_volGgroup} is taken into account.)

\subsection{Gauge group SO(3)}
In this case we start with the partition function in the sector with monopole charge $u=-1$
\begin{equation}
\widetilde{Z}(\alpha; \epsilon;-1) = \displaystyle\sum_{n=1}^{\infty} (-1)^{n}\, n^{\alpha}\,
\exp\left(-\,\epsilon\,\pi^2 n^2\right), \label{eq:12}
\end{equation}
which we rewrite as
\begin{equation}
\widetilde{Z}(\alpha; \epsilon;-1)= \frac{1}{\left(1+e^{i\pi\alpha}\right)} \displaystyle\sum_{n=-\infty}^\infty 
n^{\alpha}\, \exp\left(-\,\pi^2\epsilon\, n^2 + i\pi n\right). \label{eq:13}
\end{equation}
After a Poisson resummation, it takes the form
\begin{equation}
\widetilde{Z}(\alpha; \epsilon;-1) = \frac{1}{2\left(\pi^2 \epsilon\right)^{\frac{1+\alpha}{2}}}\,
\Gamma\left(\frac{1+\alpha}{2}\right)\, \sum_{k=-\infty}^{\infty} 
{}_1F_1\left(\frac{1+\alpha}{2},\,\frac{1}{2},\,- \frac{\left(k-\frac{1}{2}\right)^2}{\epsilon}\right).
\label{eq:14}
\end{equation}
Now, for small $\epsilon$, we use the asymptotic expansion Eq.\refb{eq:KummerAsymp} of the confluent 
hypergeometric function 
to write the $k$-th term in the series as
\begin{equation}
\frac{\sqrt{\pi}}{\Gamma\left(-\frac{\alpha}{2}\right)}   
\left(\frac{(2k-1)^2}{4\epsilon}\right)^{-\frac{1+\alpha}{2}} 
\sum_{m=0}^\infty \frac{\left(\frac{1+\alpha}{2}\right)_m\,\left(\frac{2+\alpha}{2}\right)_m}{m!} \,
\left(\frac{(2k-1)^2}{4\epsilon}\right)^{-m}, \label{eq:15}
\end{equation}
up to exponentially small terms. Summing over $k$ we get
\begin{eqnarray}
\widetilde{Z}(\alpha ; \epsilon; -1) &=& 
\frac{\Gamma\left(\frac{1+\alpha}{2}\right)}{\pi^{\frac{1}{2}+\alpha}\,\Gamma\left(-\frac{\alpha}{2}\right)}
\sum_{m=0}^\infty \frac{1}{m!}\,\left(\frac{1+\alpha}{2}\right)_m\,
\left(\frac{2+\alpha}{2}\right)_m\,\zeta\left(2m+\alpha+1,\frac{1}{2}\right)\,{\epsilon}^{m} \nonumber\\
&{}&\quad +\; \text{exponentially small terms},
\label{eq:16}
\end{eqnarray}
where $\zeta(s, q)$ is the Hurwitz zeta function. 

As anticipated, the expansion for the partition function of the SO(3) Yang-Mills theory does not have any 
negative or fractional power of $\epsilon$. The coefficients which generalize the SO(3) intersection numbers
in Eqs.\refb{eq:mj2} and \refb{coeff-poly-so3}, are
\begin{equation}
a_m = 
\frac{\Gamma\left(\frac{1+\alpha}{2}\right)\,\left(\frac{1+\alpha}{2}\right)_m
\left(\frac{2+\alpha}{2}\right)_m}{\pi^{\frac{1}{2}+\alpha}\,\Gamma\left(-\frac{\alpha}{2}\right)\,m!}\,
\zeta\left(2m+\alpha+1,\frac{1}{2}\right), \quad m=0,1,\cdots . \label{eq:16bis}
\end{equation}
We see that $\widetilde{Z}(\alpha; 0)$ is finite:
\begin{eqnarray}
a_0 \:=\: \widetilde{Z}(\alpha; 0)\:=\:
\frac{\Gamma\left(\frac{1+\alpha}{2}\right)}{\pi^{\frac{1}{2}+\alpha}\,\Gamma\left(-\frac{\alpha}{2}\right)}
\zeta\left(\alpha+1,\frac{1}{2}\right),\label{eq:16tris}
\end{eqnarray}
even though  $\mathcal{M}\left(\bar{\mathcal{S}}_{g,p}\right)$ is expected to be infinite dimensional for 
non-integer $\alpha$. We believe that in a proper mathematical set up, $a_0$ can be interpreted as an
appropriately renormalized ``symplectic volume''.  The fact that $\widetilde{Z}(\alpha; 0)$ is finite only for moduli
spaces which are smooth, is a non-trivial consistency check of our proposal, namely that the numbers  $a_m$  in 
Eqs.\refb{eq:11 bis} and \refb{eq:16bis} do indeed encode topological information of some well defined moduli 
spaces.

\subsubsection{Comparison at negative integer \texorpdfstring{$\alpha$}{}}
As in case of SU(2), the results of the previous section can be recovered by setting $\alpha=(2-2g)$ in 
Eq.\refb{eq:16}. We use the relation $\zeta\left(s,\frac{1}{2}\right) = \left(2^s - 1\right)\,\zeta(s)$ between
the Hurwitz and the Riemann zeta functions, and use the identities for the zeta and $\Gamma$-functions
to write
\begin{eqnarray}
\widetilde{Z}(\alpha; \epsilon) &=& \sum_{m=0}^{g-2} \frac{(-1)^m \pi^{2m} \left(2^{2m+3-2g}-1
\right)}{m!}\;
\zeta(2g-2m-2)\,\epsilon^m + \frac{(-1)^g\,\pi^{2g-2}}{2 (g-1)!}\,\epsilon^{g-1} \nonumber\\
&{}&\quad +\; \text{exponentially small terms}.\label{eq:17}
\end{eqnarray}
The evaluation of the $(g-1)$-th term needed a little care as it involves a cancellation between a simple
zero of the Pochhammer symbol, $(2-g)_{m}\sim (-1)^{g-1}\Gamma(g-1)(g-m-1)$ for $m=g-1$, with the 
simple pole of the zeta function, $\zeta(2m-2g-1)\sim \displaystyle\frac{1}{2(m-g+1)}$.  

\section{SU(2) theory with a holonomy along a boundary}
In the case of the SU(2) Yang-Mills on a Richards surface without boundary we found that the small
$\epsilon$ expansion of the partition function contained a term with fractional power.  We think of this 
as an indication that the moduli space of SU(2) flat connections on Richards surfaces is singular, as in 
the case of finite genus surfaces. As expected, no such singularity is found in the case of SO(3). More 
generally, we expect that the moduli space of flat SU(2) connections on a Richards surface with a fixed 
holonomy (other than identity) along its base boundary must be nonsingular, again as is the case for 
finite genus surfaces. We will confirm this expectation by carrying out an expansion of the partition 
function (of the SU(2) theory for definiteness) in small $\epsilon$ for Richards surface with a fixed 
nontrivial holonomy along its boundary. Moreover, the additional contribution from the holonomy gives 
a refined generating function, which depends on an additional parameter. 

The partition function of SU(2) Yang-Mills theory on a surface with a boundary can be thought of as a 
map from the holonomy $U_0$ at the boundary to the complex numbers. In particular, for the one 
parameter subgroup $U_0(\theta)= \exp\left(i\theta J_3 \right)$ (where $0\leq\theta<2\pi$) we have
\begin{equation}
Z(\alpha; \epsilon; \theta) = \sum_{n=1}^\infty n^{\alpha - 1}\exp\left(-\,\epsilon\,\pi^2 n^{2}\right)\;
\frac{\sin\left(n\theta/2\right)}{\sin\left(\theta/2\right)}, \label{pfwHolonomy}
\end{equation}
where $\chi_{n}(U_0(\theta)) = \displaystyle{\frac{\sin\left(n\theta/2\right)}{\sin\left(\theta/2\right)}}$ is the 
trace of $U_0(\theta)$ in the $n$-dimensional representation. Following in the steps of the previous 
sections, we write this as
\begin{equation}
Z(\alpha;\epsilon;\theta) = \frac{1}{\left(1+e^{i\pi\alpha}\right)\sin\frac{\theta}{2}} 
\sum_{n=-\infty}^\infty n^{\alpha - 1}\exp\left(-\, \epsilon\,\pi^2 n^2\right)\;\sin\left(\frac{n\theta}{2}\right),
\label{eq:18}
\end{equation}
and performing a Poisson resummation to get
\begin{eqnarray}
Z(\alpha; \epsilon;\theta) \: = \:  -\; 
\frac{2^{\alpha - 2}\Gamma\left(\frac{1+\alpha}{2}\right)}{{\epsilon}^{\frac{1+\alpha}{2}}\,\sin(\theta/2)}  
&{}& \!\!\!\!\!\!\!\! 
\sum_{k=-\infty}^\infty \left((4k\pi-\theta)\,{}_1F_1\left(\frac{1+\alpha}{2},\,\frac{3}{2},\,-\,
\frac{(4\pi k-\theta)^2}{4\epsilon}\right) \right.\nonumber\\ 
&{}& -\; \left.(4k\pi+\theta)\,{}_1F_1\left(\frac{1+\alpha}{2},\,\frac{3}{2},\,-\,
\frac{(4\pi k+\theta)^2}{4\epsilon}\right)\right). \label{eq:19}
\end{eqnarray}
Using the asymptotic expansion Eq.\refb{eq:KummerAsymp}, 
$Z(\alpha; \epsilon; \theta)$ is  
\begin{eqnarray}
-\, \frac{2^{\alpha-3}\sqrt{\pi}\,
\Gamma\left(\frac{1+\alpha}{2}\right)}{{\epsilon}^{\frac{1+\alpha}{2}}\,\sin\left(\frac{\theta}{2}\right)} 
&{}&\!\!\!\!\!\!\!\!\!\! 
\sum_{k=-\infty}^\infty \left[ \frac{(4\pi k -\theta)  \left(4\epsilon\right)^{(1+\alpha)/2}}{\left|4\pi k - 
\theta\right|)^{\alpha+1}\Gamma\left(\frac{2-\alpha}{2}\right)}
\sum_{m=0}^\infty \frac{\left(\frac{\alpha}{2}\right)_m \left(\frac{1+\alpha}{2}\right)_m}{m!}
\left(\frac{4\epsilon}{(4\pi k - \theta)^2}\right)^m\right.\nonumber \\
&{}& \left. -\,  \frac{(4\pi k +\theta)\left(4\epsilon\right)^{(1+\alpha)/2}}{\left|4\pi k + \theta\right|^{\alpha+1}
\Gamma\left(\frac{2-\alpha}{2}\right)}
\sum_{m=0}^\infty \frac{\left(\frac{\alpha}{2}\right)_m \left(\frac{1+\alpha}{2}\right)_m}{m!}
\left(\frac{4\epsilon}{(4\pi k + \theta)^2}\right)^m\right] \nonumber\\
&{}&\qquad +\; \text{exponentially small terms}
\label{eq:20}
\end{eqnarray}
which, after doing the sum over $k$, gives 
\begin{eqnarray}
Z(\alpha; \epsilon;\theta) = \, \frac{2^{1+\alpha}\sqrt{\pi}\,
\Gamma\left(\frac{1+\alpha}{2}\right)}{\sin\left(\frac{\theta}{2}\right)
\Gamma\left(-\frac{\alpha}{2}\right) }
&{}&\!\!\!\!\!\!\!\!\!\!
\sum_{m=0}^\infty \frac{\left(\frac{2+\alpha}{2}\right)_{m-1} 
\left(\frac{1+\alpha}{2}\right)_m}{(2\pi)^{2m+\alpha}\; m!} \label{eq:21}\\
&{}&\times\;\left[\zeta\left(2m+\alpha,1-\frac{\theta}{4\pi}\right) - 
\zeta\left(2m+\alpha,\frac{\theta}{4\pi}\right)\right]{\epsilon}^m. \nonumber
\end{eqnarray}
As expected, this expression does not have any fractional power of $\epsilon$.

\section{Conclusions}\label{sec:concl}
The moduli spaces of flat connections on two dimensional surfaces are of considerable importance. 
Consequently, these have been studied extensively by mathematicians as well as physicists using 
diverse tools. In particular, the partition function of 2dYM on surfaces contains information about 
various intersection numbers of the moduli spaces of flat connections on principal bundles on these 
surfaces. 

In this paper we made a step towards extending these structures to surfaces of infinite genus. More
precisely, we analyzed the weak coupling expansion of  the partition function of 2dYM on a class of 
such surfaces, which we call Richards surfaces, that have a recursive structure. The partition function 
of 2dYM on Richards surfaces was computed in \cite{Kumar:2014jpa}. Here we analyzed in detail its weak 
coupling expansion for the gauge groups SU(2) and SO(3), with and without the insertion of a Wilson 
line.

We were able to do this by considering the partition function as an {\it analytic} function of the complexified 
Euler characteristic of a surface.  Considering the Euler characteristic at fractional values as appropriate to 
Richards surfaces (or continuing analytically to negative integral values as appropriate for surfaces of
finite characteristic), we found that the corresponding partition function can be split into three contributions. 
First, there is a term which is finite whenever the corresponding theory admits a {\it smooth} moduli space of 
classical solutions at zero coupling. It seems reasonable to interpret this as the symplectic volume of the 
(infinite dimensional) moduli space of flat connections on Richards surfaces. The fact that it turns out to be 
finite whenever the moduli space is smooth is a consistency check that this geometrical interpretation is 
coherent.

Secondly, there is  a contribution to the partition function that is perturbative (in $\epsilon$). In the finite 
genus case the perturbative series stops at a finite order. It was understood by Witten\cite{Witten:1992xu}
that the finite number of perturbative coefficients are the intersection numbers of certain cohomology classes 
on the (finite-dimensional) moduli space of flat connections on surfaces of finite genus. In the Richards case 
we discovered that the perturbative expansion never stops. This appears to be consistent with the expected
infinite dimensionality of the corresponding moduli space of flat connections. The infinite number of coefficients 
of this expansion, for which we gave explicit formulas, are therefore the natural candidates  for the intersection 
numbers on some appropriately defined moduli space of flat connections over Richards surfaces. We hope to 
report on this geometric interpretation in the near future.

Finally, for both the finite and the infinite genus cases, there are non-perturbative, instanton-like corrections
which are exponentially small. Although these can be extracted from the expressions we have derived, we 
have not analyzed these terms in details to connect them to non-flat classical solutions. This exercise, which
we leave for the future, may be a rewarding one since it could provide a proper field theoretic context in which 
to extend the tools and concepts of  resurgent analysis which has so far been applied largely to quantum 
mechanical systems.

\bigskip

\noindent{\bf Acknowledgments:} It is a pleasure to thank Fran\c{c}ois Labourie for useful discussions. 
The work of DK is supported by a research fellowship from the Council of Scientific Research (CSIR), 
India, and that of CI, in part, by the INFN, Italy. DG gratefully acknowledges the hospitality at the 
Universit\`{a} degli studi di Genova, where this project was initiated during a visit that was supported 
by the SERP-Chem Master and the Erasmus Mundus programme of the European Union.
 



\end{document}